\documentstyle[multicol,epsfig,prb,aps]{revtex}

\ifpreprintsty
\def\multb{ }
\def\multe{ }
 \else
\def\multb{ \begin{multicols}{2}}
\def\multe{ \end{multicols}}
 \fi

\begin{document}   
\draft
\title{Electronic structure of MgB$_2$: x-ray emission and absorption studies}
\author{E.Z. Kurmaev$^1$, I.I. Lyakhovskaya$^2$, J. Kortus$^3$, 
A. Moewes$^4$,
N. Miyata$^5$, M. Demeter$^6$, M. Neumann$^6$, M. Yanagihara$^5$, 
M. Watanabe$^5$, T. Muranaka$^7$ and J. Akimitsu$^7$}
\address{%
$^1$Institute of Metal Physics, 
    Russian Academy of Sciences-Ural Division, 620219 Yekaterinburg, Russia\\
$^2$Institute of Physics, St. Petersburg State University,
    198504  St. Petersburg, Russia\\
$^3$Max-Planck-Institut f{\"u}r Festk{\"o}rperforschung,
    D-70569 Stuttgart, Germany\\
$^4$Department of Physics and Engineering Physics, University of Saskatchewan, 
    116 Science Place Saskatoon, Saskatchewan S7N5E2, Canada\\
$^4$Research Institute for Scientific Measurements, 
    Tohoku University, Sendai 980-8577, Japan\\
$^5$University of Osnabr{"u}ck, Sektion of Physik, D-49069 Osnabr{\"u}ck, Germany\\
$^6$Department of Physics, Aoyama-Gakuin University, Tokyo 157-8572, Japan}
\date{\today}
\maketitle
\begin{abstract}  
Measurements of x-ray emission and absorption spectra of the constituents of 
MgB$_2$ are presented. The results obtained are in good agreement with 
calculated x-ray spectra, with dipole matrix elements taken into account.
The comparison of x-ray emission spectra of graphite, AlB$_2$, and 
MgB$_2$ in the binding energy scale supports the idea of charge transfer 
from $\sigma$ to $\pi$ bands, which creates holes at the top of the 
bonding  $\sigma$  bands and drives the high-T$_c$ 
superconductivity in MgB$_2$.
\end{abstract}
\multb
\section{Introduction}
The recent discovery of superconductivity in MgB$_2$ with a $T_c$ close to
40 K \cite{ref1} was an unexpected experimental achievement.
Up to now, it is the highest T$_c$ value reported for any binary compound 
\cite{ref2}. This value of $T_c$ is much higher than 
previously expected within the context of BCS theory \cite{ref3}. 
Not surprisingly, its discovery has given rise to much experimental 
and theoretical activity, having
raised the possibility of a whole new class of superconductors.

The experimentally observed B isotope shift of T$_c$ \cite{ref4} 
and other experimental data \cite{ref5} suggest 
conventional BCS $s$-wave electron-phonon coupling. 
However, careful analysis of the temperature and magnetic field
dependence of the specific heat \cite{heat} suggests anisotropic or
multiple gaps. The reported values of $2\Delta/k_B T_c= 1.2-4$ from
tunneling measurements \cite{tunnel} also raise the possibility 
of multiple gaps, although the values below the BCS weak coupling
limit of 3.5 have been attributed to surface effects.
Even though there is growing evidence for
conventional BCS $s$-wave electron-phonon coupling, the experimental
picture is not yet entirely clear.
In addition to theoretical explanations based on 
BCS theory, \cite{ref6,ref7,ref8} an alternative explanation based on 
hole superconductivity has been proposed.\cite{ref9}
Both theories are based on the results of band structure calculations 
of MgB$_2$. 

X-ray emission and absorption spectroscopies are powerful probes of 
the electronic structure of solids. 
Photon emission and absorption involve a transistion between electronic
states.  In the soft x-ray
regime, one of the states is a localized, dispersionless core level. 
This allows for the interpretion of the measured spectra in terms of 
unoccupied states for absorption and occupied states for emission. 
Since dipole selection rules govern the transitions to or from the core 
level, it is actually the angular-momentum-resolved density of states (DOS)
that is measured. Furthermore, since the core level 
is associated with a specific element in the compound, 
x-ray absorption and emission are also element specific. Finally, they
have the advantage of being relatively insensitive to the quality of the
sample surface, unlike 
x-ray photoelectron spectroscopy (XPS) or
ultraviolet photoemission, where in order to measure the bulk 
electronic structure it is necessary to prepare atomically clean, 
stoichiometric, and ordered surfaces, which are impossible to realize 
for sintered samples such as MgB$_2$.

Recently, a high-resolution photoemission study of a sintered powder 
sample of MgB$_2$ was carried out \cite{ref10}, 
however the behavior of the spectral function was analyzed only in 
the vicinity of the Fermi level. 
In the present paper, the x-ray emission and absorption spectra 
(XES and XAS) of the constituents 
have been studied in MgB$_2$ and related 
compounds (graphite and AlB$_2$). 
The results obtained are compared with the partial density 
of states and first-principles calculations of the 
intensities of x-ray spectra which 
take dipole matrix elements and selection rules into account.

\section{Experimental Details}
Both pressed powder and sintered polycrystalline MgB$_2$ samples were 
used for measurements of the XES.
The sintered polycrystalline sample was prepared as described in 
Ref.\ \onlinecite{ref1}. 
X-ray diffraction measurements show that the sample is single-phased 
and electrical resistivity and DC magnetization measurements confirm the onset
of a sharp superconducting transition at 39.5 K. 
The B K-emission and absorption spectra were studied on Beamline 8.0.1 at the
Advanced Light Source (ALS) at Lawrence Berkeley National Laboratory
employing the soft X-ray fluorescence endstation.\cite{Moewes}
Emitted radiation was measured using a Rowland circle type spectrometer 
with spherical gratings and a multichannel two-dimensional detector. 
The measurements of the Mg L-emission spectra were performed using an 
ultrasoft x-ray grating spectrometer (R=1 m, n=600 l/mm) 
with electron excitation.\cite{ref11}  
The B K$\alpha$ and Mg L$_{2,3}$ XES were measured with excitation energies 
far from the B $1s$ and Mg $2p$ thresholds (non-resonant
spectra), with an energy resolution of 0.3 eV.

The Mg 2$p$ absorption spectra were also measured at 
the beamline BL-12A at the Photon Factory in KEK using photons
from a synchrotron source. 
The energy resolution near the Mg 2$p$ threshold (50 eV) is 0.5 eV,
using a 0.1 mm monochromator slit-width. 
For Mg 2$p$ measurements, we used a Si filter despite the Si 2$p$ absorption 
threshold at about 100 eV because at this beamline non-negligible 2nd 
order light is included in this energy region 
The absorption spectra were taken by recording the total electron yield (TEY) 
sample drain current. 
To remove surface contamination before the measurements, the sample 
was scraped with sandpaper and then striped off with vinyl tape repeatedly 
until the mark left on the tape was uniform. 
The vacuum was below $1.0\times 10^{-6}$ torr
and the measurements were carried out at room temperature.

In order
to determine the position of the Fermi level and convert x-ray spectra 
to the binding energy scale of XPS (difference of measured XES energies
and a selected XPS core level energy),
B 1$s$ and Mg 2$p$ core levels were measured. 
As mentioned above, XPS for valence band states is very
sensitive to surface contamination, nevertheless the binding energies of
core levels can be determined after cleaning the surface.
The XPS measurements have been carried out with an ESCA 
spectrometer manufactured by Physical Electronics (PHI 5600 ci). 
The monochromatized Al K$\alpha$ radiation had an FWHM of 0.3 eV 
and combined with the energy resolution of the 
analyzer (1.5\% of the pass energy) results in an estimated 
energy resolution of somewhat less than 0.35 eV. 
XPS measurements of a MgB$_2$ sample fractured in high vacuum 
have shown less oxygen content on the surface than those 
of sintered material. Further reduction of the oxygen 
content was achieved by ion etching. After cleaning the surface, 
we obtained the following values for the binding energies associated with the 
core levels: B $1s$ (185.5 eV) and Mg $2p$ (49.5 eV).
These results agree well with recent XPS studies of MgB$_2$
(the  Mg $2p$ core level energy agrees within 0.2 eV) \cite{XPS}
which shows that the XPS core levels are not significantly influenced 
by oxidized surfaces, as shown in this study by comparing as-grown
and etched surfaces in which the oxidized layer is effectively removed. 
This supports the use of these values in Fig.\ \ref{fig4}
to convert XES spectra to the binding 
energy scale.

\section{Results and Discussion}
The states at the Fermi level derive primarily from 
B and so the resulting 
band structure can be understood in terms of the boron sublattice. 
Mg can be described as ionized (Mg$^{2+}$) in this compound.
However, the electrons donated to the system are not localized on the anions, 
but rather are distributed over the whole crystal. 
The $\sigma$ ($sp^2$) bonding states which have B p$_{xy}$ character
are unfilled, in contrast to graphite where the $\sigma$ states 
are completely filled and form strong covalent bonds.
These $\sigma$ states form two small cylindrical Fermi surfaces around the
$\Gamma$--A line and the holes at the top of the 
bonding $\sigma$ bands are believed to couple strongly to optical
B--B modes and to play a key role in the superconductivity
of MgB$_2$.

Using the full potential LAPW code WIEN97, \cite{ref12} we calculated 
the near edge absorption and emission spectra. 
According to the final-state rule formulated by von Barth and Grossmann, 
\cite{Barth}
accurate XES and XAS of simple metals may be obtained from 
ordinary one-electron theory if the relevant dipole matrix
elements are calculated from valence functions obtained in the potential
of the final state of the x-ray process: in other words, a potential
reflecting the fully screened core hole for absorption but not for emission.
Because we neglected core relaxation effects in our calculations,
we expect significantly better agreement between theory and experiment
for emission spectra. 
The calculated spectra are Lorentz broadened with a 
spectrometer broadening of 0.35 eV, with additional lifetime broadening 
for emission spectra. 

\begin{figure}      
\begin{center}
\epsfig{file=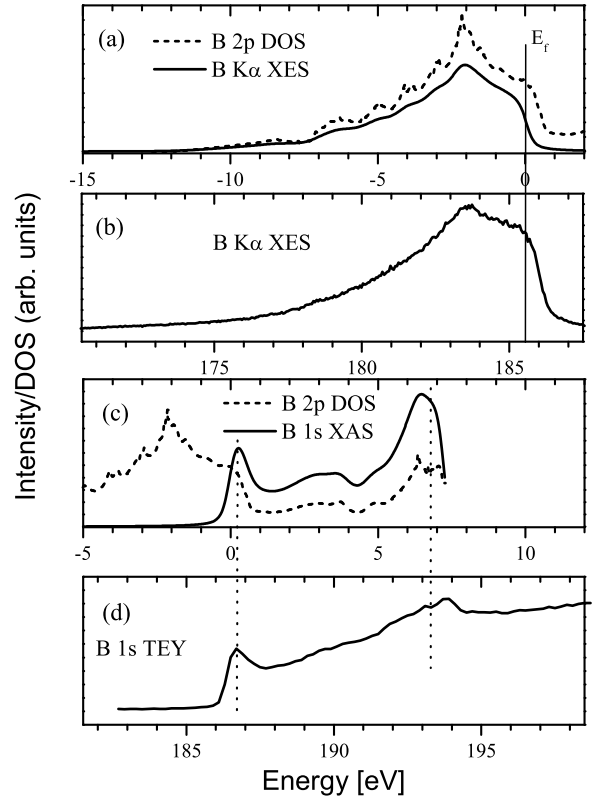,width=\linewidth}
\end{center}
\caption{Calculated (a, c) and experimental (b, d) B K emission 
and absorption spectra of MgB$_2$.
The emission spectrum was obtained from electron excitation far from
resonance and is an accumulation of several scans. The absorption spectrum was 
recorded from the total electron yield (TEY) sample drain current using
photons from a synchrotron source.
} 
\label{fig1}
\end{figure}

\begin{figure}
\begin{center}
\epsfig{file=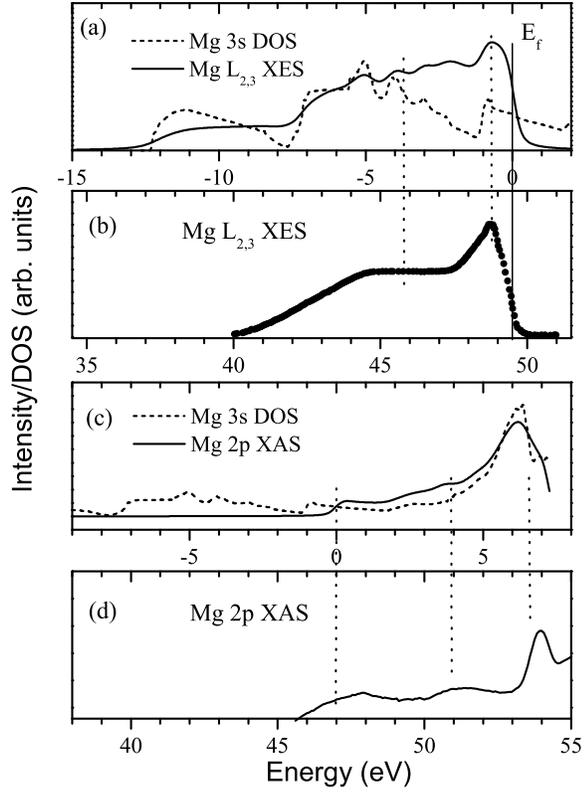,width=\linewidth,clip=true}
\end{center}
\caption{Calculated (a, c)  and experimental (b, d) Mg L emission and 
absorption spectra of MgB$_2$. The density of states is 
significantly changed by the dipole matrix elements, which are
necessary in order to recover the experimental XES shape.
 }
\label{fig2}
\end{figure}

\begin{figure}
\begin{center}
\epsfig{file=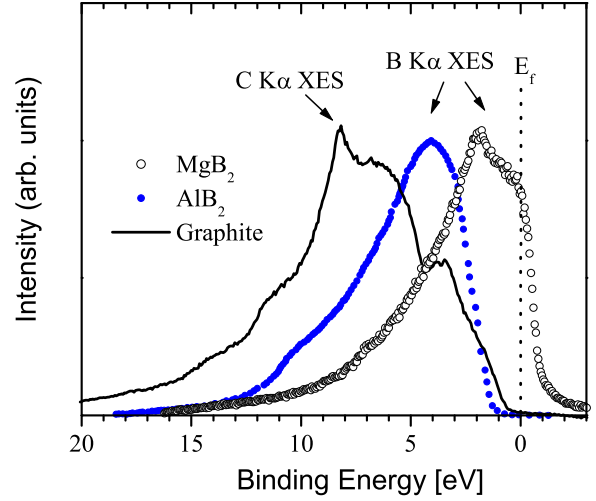,width=\linewidth,clip=true}
\end{center}
\caption{Comparison of x-ray emission spectra of hexagonal graphite, 
AlB$_2$, and MgB$_2$ using the binding energy scale (difference of
XES energies and the $2p$ core level energy obtained from XPS).
The major maximum originates from $\sigma$ ($sp^2$) states in all
materials. The observed shift from graphite to MgB$_2$ supports
the theoretical results of Ref.\ \protect\onlinecite{ref7}.} 
\label{fig4}
\end{figure}

Theoretical B K emission and absorption spectra of MgB$_2$, which 
according to dipole selection rules ($1s \rightarrow 2p$ transition) 
probe B 2$p$ occupied and unoccupied states, respectively, 
are presented in Fig.\ \ref{fig1} (a, c). These calculations show 
that the intensity distributions of B K emission and absorption follow
the B 2$p$ partial density of states very closely because the radial 
dipole matrix elements are monotonically increasing functions of energy 
within the valence and conduction bands. 
The experimental B K emission and absorption spectra (Fig.\ \ref{fig1} (b, d)) 
are in good agreement with the calculated spectra.
Note that the position of the Fermi level of XAS
cannot be determined with the help of XPS measurements of core levels
due to the different final states. The experimental and calculated
XAS spectra shown in Figs.\ \ref{fig1}--\ref{fig2} 
are compared by aligning the Fermi levels.
Our results  are also in good agreement with other recent 
experimental studies.\cite{Callcott,Nakamura}
Callcott {\it et al.} \cite{Callcott} have reported soft x-ray fluorescence
measurements and also XAS for the $K$-edge of B in MgB$_2$.
Comparing their results to our TEY  (Fig.\ \ref{fig1}d) we
find similar structures around 187 eV and 193 eV as observed
in Ref.\ \onlinecite{Callcott}. Similar results were also obtained by Nakamura
{\it et al.} \cite{Nakamura}. 
We do not find the peak labeled C in Ref.\ \onlinecite{Callcott} around 195 eV,
which was associated with boron oxide.
We also obtained good agreement 
comparing our XES results with other B K-emission 
experimental data.\cite{Callcott,Nakamura} 

Calculated Mg L emission and absorption spectra which probe occupied and 
unoccupied Mg 3$s$ states are shown in Fig.\ \ref{fig2} (a, c). 
The intensity distribution associated with
the Mg L emission differs somewhat from the
Mg 3$s$ partial DOS.
The DOS is significantly changed by the dipole matrix elements
which are necessary in order to recover the experimental XES shape.
The contribution to the x-ray intensity is larger for states near 
the Fermi level relative to those at the bottom of the valence band, 
in accordance with the energy dependence of the radial dipole 
matrix elements. 
Again, we note reasonable agreement between the 
calculated (Fig.\ \ref{fig2} (a, c)) 
and experimental spectra (Fig.\ \ref{fig2} (b, d)).

All density-functional calculations appear to agree that Mg is substantially
ionized \cite{ref6,ref7,ref8}. In order to provide direct
experimental evidence for charge transfer from Mg, we compared
the Mg L emission spectrum in MgB$_2$ to that in
pure Mg metal. The shift of the Mg core $2p$-level with
respect to the Fermi level contains this information and will
be called chemical shift. This does not involve counting of
electrons in order to obtain the chemical potential, which
would depend critically on the shape of the bands.
Normally, one would expect a core level shift
towards higher binding energies (positive shift) in losing valence charge 
because less electrons screen the Coulomb potential weaker and therefore
the electrons are bound stronger.

Band structure calculations reveal, the Mg $2p$ level
shows significant hybridization with the B $p_z$ level, which raises 
the former while lowering the latter, 
in difference to the simple picture outlined above.
The mechanism is similar to an interacting two-level system,
where new states are formed, one level is lowered in energy
and the other level is pushed up in energy proportional to the
overlap (hybridization) between the states.

A negative chemical shift of about 0.5 eV is found in the MgB$_2$ 
spectrum with respect to that of pure Mg, which we hold as evidence for charge 
transfer from Mg to B atoms in this compound. 
The same negative chemical shift of 0.5 eV has been observed recently in XPS 
measurements \cite{XPS} as well.

This is a very important
effect because it lowers the $\pi$ ($p_z$) bands relative to the bonding
$\sigma$ ($sp^2$) bands. This lowering of the B $\pi$ bands 
relative to the $\sigma$ bands, compared to graphite,  
causes $\sigma \rightarrow \pi$ charge transfer      
and $\sigma$-band hole doping, 
driving the superconductivity in MgB$_2$ \cite{ref7}. 
To investigate this prediction further, we compared the
K$\alpha$ XES ($2p\rightarrow 1s$ transition) 
for graphite, AlB$_2$, and MgB$_2$
by alignment of the Fermi levels which were determined
 in the binding energy scale 
using XPS measurements of core levels: B $1s$ (MgB$_2$)=185.5 eV, B $1s$ 
(AlB$_2$)=188.5 eV, \cite{ref15} and C $1s$ (graphite)=284.5 eV \cite{ref15} 
(Fig.\ \ref{fig4}).
As the figure shows, the major maximum originating from 
$\sigma$ states is shifted in MgB$_2$ towards the 
Fermi level with respect to that of graphite.
AlB$_2$ occupies an intermediate position due to the higher electron 
concentration compared to MgB$_2$ which results in filling of 
the $\sigma$ bands, 
decreasing $N(E_f)$ and finally destroying superconductivity.

\section{Conclusion}
In conclusion, we have measured x-ray emission and absorption spectra of 
the constituents of the new superconductor MgB$_2$ and found good 
agreement with results of band structure calculations and 
in particular calculations of intensities of x-ray spectra 
taking the necessary matrix elements into account. 
Further, according to our findings magnesium is positively charged in this 
compound, which supports the results of electronic structure calculations. 
The comparison of x-ray emission spectra of graphite, AlB$_2$, and MgB$_2$ 
supports the idea of superconductivity driven by hole doping of 
the covalent $\sigma$ bands. 
While the experimental results of our study cannot give direct insight into the
mechanism of the superconductivity, they do support and lend credence 
to the standard band structure methods used in the theoretical 
analysis of this new and exciting material. This information could
prove important in understanding and answering the questions
which still exist. 

\acknowledgements  
We thank I. I. Mazin and O. Gunnarson for helpful 
discussions and comments
and J. E. Pask for a critical reading of the manuscript.
The Russian State Program on Superconductivity, Russian Foundation for Basic 
Research (Project 00-15-96575) and NATO Collaborative Linkage Grant 
supported this work. JK would like to thank the Schloe{\ss}mann Foundation 
for financial support. Funding by the President's NSERC
fund of the University of Saskatchewan is gratefully acknowledged.

\multe
\end{document}